\documentclass[aps,prl,twocolumn]{revtex4}

\usepackage{amsmath}
\usepackage{amssymb}
\usepackage{amsthm}
\usepackage{psfrag}
\usepackage{graphicx}

\newcommand{\be}{\begin{equation}}
\newcommand{\ee}{\end{equation}}
\newcommand{\beqa}{\begin{eqnarray}}
\newcommand{\eeqa}{\end{eqnarray}}

\def\e{{\rm e}}
\def\d{\partial}
\newcommand{\sm}[1]{{\scriptscriptstyle \rm #1}}
\newcommand{\vev}[1]{\langle #1 \rangle}
\newcommand{\lp}{\left}
\newcommand{\rp}{\right}
\newcommand{\eff}{{\rm eff}}
\newcommand{\GL}{{\scriptscriptstyle\rm GL}}
\newcommand{\OO}{{\cal O}}

\newcommand{\LL}{{\cal L}}
\newcommand{\NN}{{\cal N}}

\newcommand{\eg}{{\em e.g.}}
\newcommand{\wt}{\widetilde}
\newcommand{\ds}{\displaystyle}

\input epsf%
\begin{document}
\title{Flux Periodicities and Quantum Hair on Holographic Superconductors}
\author{Marc Montull,\!$^a$ Oriol Pujol\`as,\!$^a$
  Alberto Salvio$^{a,b}$ and Pedro J. Silva,\!$^{a,c}$}
  \affiliation{{$^a$}
 \it Departament de F\'isica and IFAE, Universitat Aut\`onoma de Barcelona,
 Bellaterra 08193 Barcelona Spain}
 \affiliation{{$^b$}\it Scuola Normale Superiore and INFN, Piazza dei Cavalieri 7, 56126 Pisa, Italy}
 \affiliation{{$^c$} \it 
ICE-CSIC, Universitat Aut\`onoma de Barcelona, 08193 Bellaterra, Barcelona Spain
}

\begin{abstract}
Superconductors in a cylindrical geometry  
respond periodically to a cylinder-threading magnetic flux, 
with the period changing from $hc/2e$ to $hc/e$ 
depending on whether the Aharonov-Bohm effects are suppressed or not. 
We show that Holographic Superconductors present a similar phenomenon, 
and that the different periodicities follow from classical no-hair theorems.
We also give the Ginzburg-Landau description of the period-doubling phenomenon.
~\\[-5mm]
\end{abstract}

\maketitle

\thispagestyle{plain}%

{\em Introduction:}
The study of superconductors (SCs) in multiply connected geometries 
has provided important applications 
as well as theoretical insights. 
The simplest configuration consists of a superconducting  hollow cylinder 
of negligible width threaded by a magnetic flux $\Phi_B$. 
Two classic effects in this setup are
i)
the Aharonov-Bohm (AB) interference between charged quanta 
winding around the cylinder. This phenomenon exhibits a pe\-rio\-dic dependence 
(of, say, the partition function) on the flux $\Phi_B$ with a period given by the 
flux quantum  $\Phi_0 = h/e$ 
\footnote{We use units where the speed of light is $c=1$.} 
where $e$ is the fundamental charge unit;
ii) 
the fact that thermodynamic quantities such as the critical temperature $T_c$ 
acquire a periodicity in the flux with period 
$\Delta \Phi_B = \Phi_0 / 2 $  \cite{LP},
which represents  a hallmark of pairing in the SC.
For short, we shall refer to this as the `Little-Parks' (LP) period.
Recently, considerable theoretical work has discussed how the LP periodicity  should
return to the quantum bound $\Phi_0$ for cylinder radius $R$ 
comparable to the coherence length $\xi_0$ \cite{Vakaryuk,Loder,WG}. 
An appropriate  statement of the LP result, then, 
is that the two $h/2e$ sub-periods are actually {\em degenerate} for $R\gg\xi_0$. 
In this Letter, we describe the LP {\em degeneracy} and 
breaking thereof 
occurring in Holographic Superconductors (HSC) \cite{HHH}. 
As we shall see, 
the dual LP effect emerges from  Black Hole no-hair theorems.

{\em Discrete Gauge Symmetries.}
Superconductivity is perhaps the simplest
realization of a {\em discrete gauge symmetry}. Indeed, in a system 
with two fields $\xi$ and $\phi$ (either fundamental or composite) with $U(1)$ gauge charges $e$ and $g\equiv N\,e$ respectively with $N$ an integer, whenever the high charge field $\phi$ condenses, then a $Z_N$ gauge subgroup is unbroken and realized nontrivially by $\xi$. 
In pairing-based SCs, $N=2$ and the 
discrete  charge amounts to the total number of electrons in the SC being even or odd. 
This charge is in principle measurable  
\eg\, 
by AB interference of 
magnetic flux tubes lassoing the SC \cite{axionHair,KW,CPW}.

Once the SC presents a nontrivial topology, a complete description of the 
system may require  non-local gauge invariant objects, 
such as the Wilson line 
$W\equiv \exp\lp({e i \oint dx^\mu a_\mu /\hbar}\rp)$
where $a_\mu$ is the gauge potential 
and the integral runs over a non-contractible circle.
For a cylinder of radius $R$
the integral in the circular direction $\chi$ gives
$W(a_\chi)=\exp\lp(2\pi e i R a_\chi/\hbar\rp) $,
a measure of the magnetic flux within the cylinder, 
$\exp\lp({2\pi i \Phi_B/\Phi_0}\rp)$. 
Another gauge invariant non-local object is the winding number (or `fluxoid')
${\cal M}\equiv{\oint dx^\mu \partial_\mu \theta}/2\pi$
where $\theta$ is the phase of the condensing field, $\phi=\rho\,e^{i\theta}$, 
the integral taken on the same path. 
${\cal M}$
parametrizes the  momentum carried  by the condensate in the $\chi$ direction
and  
satisfies the quantization condition 
${\cal M}=m\in\mathbb{Z}$, stemming from the single-valuedness of $\phi$'s `wavefunction'.
This winding number is well defined 
whenever the $U(1)$ is Higgsed. 
If  a $Z_N$ group is left unbroken then $m$ is effectively defined modulo $N$. 
Therefore, in general there are $N$ distinct 
fluxoid configurations $\phi =\rho \,\e^{i \,{m \chi / R} }$ for $\rho\neq0$. 
The LP period implies that the $N$ fluxoids are degenerate 
for large enough $R$.
%

\begin{figure}[t]
 	\[
	\begin{array}{ccc}
	    \includegraphics[width=2.2cm,height=2.2cm]{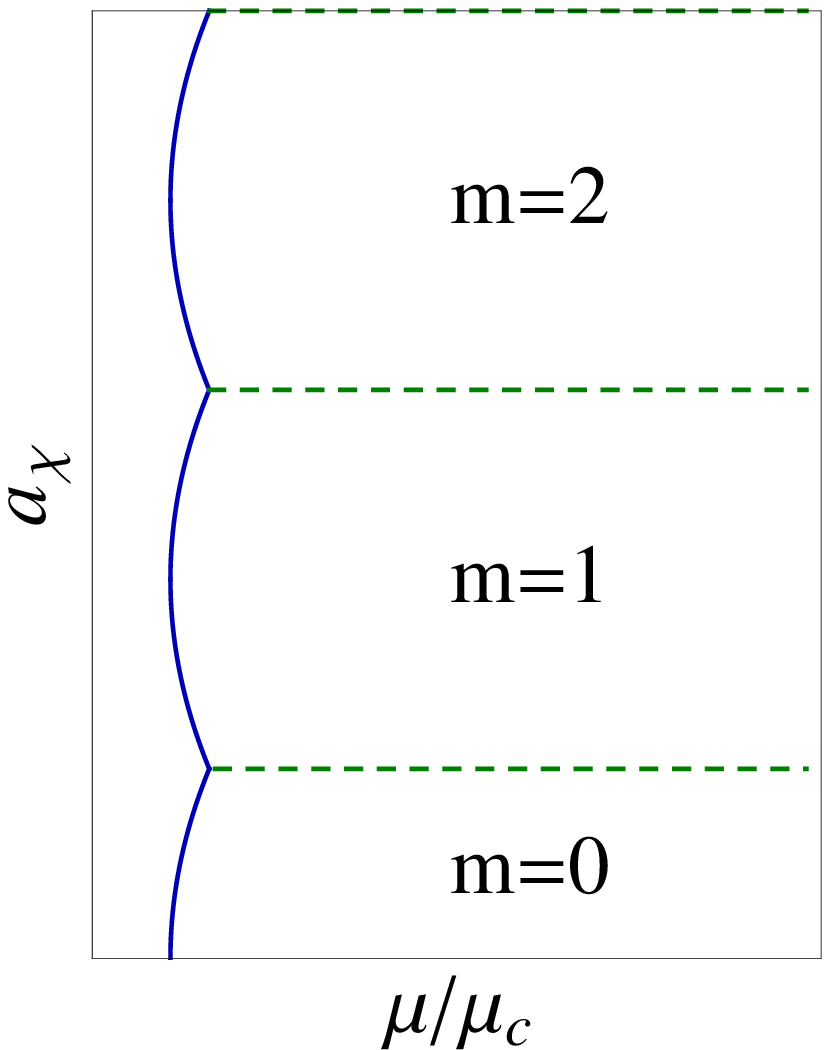}~~~~
   &   \includegraphics[width=2.2cm,height=2.2cm]{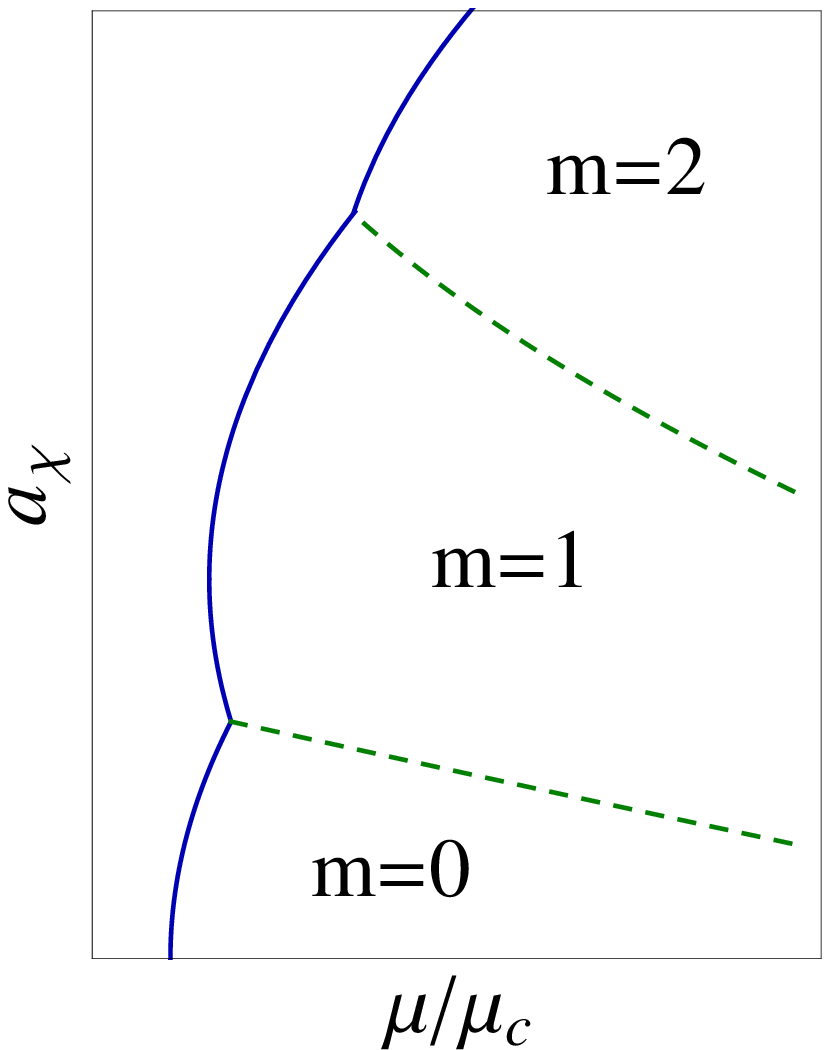}~~~~
   &   \includegraphics[width=2.2cm,height=2.2cm]{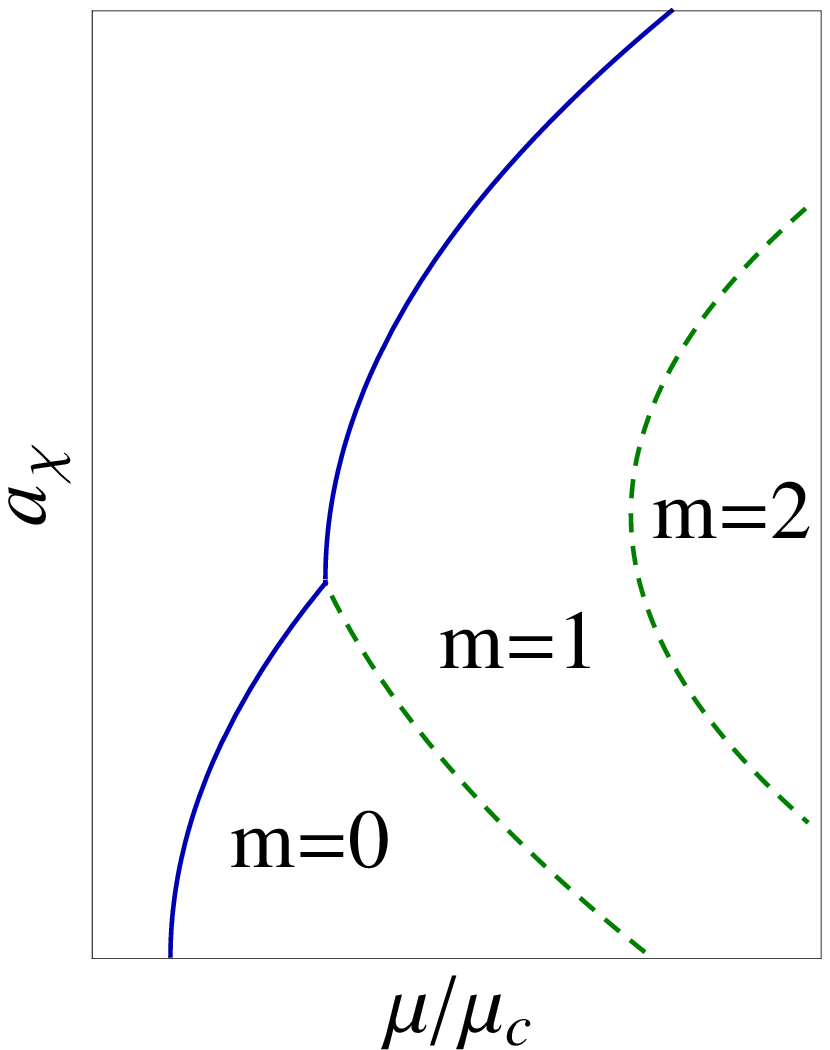}\\
   \sm{(a)}&\sm{(b)}&\sm{(c)}\\[-6mm]
	\end{array}
	\]
  \caption{{\footnotesize   $\mu$-$a_\chi$ phase diagrams 
  from a GL theory where  $M$, $b$ are (a) constant, and (b,c) increasingly more dependent on $m,W$. 
  Solid blue lines mark  SC/normal  transitions. 
  Dashed green lines mark transitions between different $m$  channels.
  For illustration, we consider a generic $N>2$ and show the first few  $m$ sectors.\\[-6mm]  
  }}\label{fig:GL}
~\\[-5mm]
\end{figure}

{\em (Uplifting) The LP Degeneracy in Ginzburg-Landau.}
Let us now  review the LP effect in terms 
of the Ginzburg-Landau (GL)  theory. 
The behaviour of the GL order parameter $\phi_\GL$ in thermodynamic equilibrium
is obtained
by minimizing an effective Lagrangian, which
close enough to the transition 
is 
approximated by
\be\label{LGL}
\LL[\phi_\GL]=|D_\mu\phi_\GL|^2-M^{2}|\phi_\GL|^2-{b}|\phi_\GL|^4
\ee
with $D_\mu\phi_\GL = (\d_\mu - i g\,a_\mu)\phi_\GL$ (hereafter $\hbar=1$),
we omit the kinetic term for $a_\mu$ and 
$M$,$b$ are constants.
%
Treating  $a_t$ as an external chemical potential $\mu$  and 
including 
the external 
flux by the minimal coupling, 
one sees that
the effective mass-squared of  $\phi_\GL$ on the fluxoid state $m$ is
$
M_\eff^2 = M^2 +  g^2 [ -\mu^2 + \big(a_\chi - {m/ gR}\big)^2  ].
$
Provided the chemical potential is large enough then $M_\eff^2<0$ and $\phi_\GL$ 
condenses. 
Since the different $m$ sectors have identical properties except for a discrete 
shift of $a_\chi$, the condensate forms in consecutive fluxoid channels
and the periodicity 
of the 
SC phase transition  is granted, 
with period 
$\Delta a_\chi  = 1/gR$ (the equivalent of the LP period $\Phi_0/2$). 
This simple picture directly leads to  the array of 
parabolas in the $\mu$-$a_\chi$ phase diagram typical of the LP effect, see Fig.~\ref{fig:GL} (a).
One realizes that the LP effect holds because $a_\chi$~and~$m$ enter in 
$\LL$ only via covariant derivatives, 
$|D_\chi \phi_\GL| = \rho|m/R-g a_\chi|$. 
Hence, a key ingredient behind the LP degeneracy is the {\em locality} of 
$\LL$ with respect to $a_\chi$ and $m$
\footnote{The $1/gR$ period persists for $\LL$ beyond the GL form \eqref{LGL}
provided only local gauge invariant operators enter.}. 

At quantum level, though, one does {\em not} expect $a_\chi$
to enter in $\LL$ via local operators only:
nothing prevents \eg\,  
the coefficients $M$ and $b$ in \eqref{LGL} to acquire a direct 
dependence on $W$ and $m$ (which we shall call `non-local' dependence for short).
A related and more familiar phenomenon is the Casimir effect -- the 
dependence  on $R$ of the vacuum energy (a tacit $\phi_\GL$-independent additive term in  \eqref{LGL}). 
In the presence of a cylinder-threading flux, the vacuum energy  depends 
on $W$ as well \cite{Hosotani}, giving  an `Aharonov-Bohm' Casimir effect.
Similarly, one  expects $M$ and $b$ to acquire  $W$-dependence. 
By locality, this must be suppressed by $R$: `AB' contributions to $M$ 
must scale like $1/R$  times a function of $W$ (and $m$, see below). 
Since the coherence length $\xi_0$ plays the role of $M^{-1}$ of the pair for $R\to\infty$, 
the AB-like effects can only be sizeable for $R$ close to $\xi_0$. 
The physical effect leading to  $m$-de\-pen\-den\-ce is  subtler. 
It seems to require a coupling~in the  La\-gran\-gian 
between the fundamental charges and the condensate's phase $\theta$   
such as $\LL_{j-\theta}\equiv j^\mu \,\partial_\mu \theta$,
which is gau\-ge-invariant if $j_\mu$ is a conserved current (say of the elec\-trons). 
This coupling is allowed in a Higgs phase and 
yields another AB-like effect if $\theta$
has winding $m\neq0$ 
\cite{Gia}.

We shall not attempt here to obtain the $W$,$m$-dependence from first principles 
(see \eg\,\cite{Vakaryuk,Loder,WG} for microscopic derivations of equivalent effects).
Instead, let us discuss its impact on the phase transition. 
It is instructive to consider a local GL model with constant $M, b$ and gradually turn on a non-local dependence on $W, m$. We illustrate the resulting $\mu$-$a_\chi$ phase diagrams in Fig~\ref{fig:GL}. Starting from a representative local model in (a), introducing a non-local dependence deforms the phase diagram in various ways:
the SC/normal critical lines for different $m$ channels become non-degenerate,
and the transitions between different fluxoid sectors bend around. 
Thus, the non-local dependence 
uplifts the degeneracy amongst different fluxoid sectors. 
The order parameter  is also affected in an interesting way. 
It is easy to show even beyond the GL approximation that if  parity is preserved and 
the coefficients in the Lagrangian $M, b, ...$ are allowed to depend on $W$ but not on $m$, then
the order parameter
$|\phi_\GL|$ is continuous across the transitions between different $m$-domains.
Conversely, a discontinuous $|\phi_\GL|$ in these transitions is a  signature of $m$-dependence 
(and hence of the coupling $\LL_{j-\theta}$) on quite general grounds.
As we shall see, the confining HSC exhibits 
this feature.

{\em Holographic Superconductors.}
The gauge/gravity duality 
relates
strongly coupled gauge theories with gravity in higher dimensions, and  has become a powerful method to study strongly interacting systems. 
To apply it to SCs \cite{HHH} one devises 
a 2+1 Conformal Field Theory (CFT) with a large number of 
`colours' $\NN$ and a  $U(1)$ symmetry. 
Appropriate quantization \cite{emergent} leads to an emergent gauge 
boson$-$  a gauged $U(1)$. Superconduc\-tivity  occurs if 
a $U(1)$-charged operator $\OO$ condenses. 
This is dual to 3+1 Anti-de-Sitter (AdS) gravity with a gauge  
and a charged scalar field (dual to $\OO$).
Fixing the che\-mical potential $\mu$, 
at high  temperature the ground state is the AdS Reissner-Nordstrom charged Black Brane (BB), which is dual to a conductor. 
Instead, at low temperature the BB develops scalar hair, 
and thus becomes dual to a SC \cite{HHH}.

An interesting development of this model consists in compactifying one spatial dimension,  corresponding to the cylindrical geometry considered here. 
With appropriate boundary conditions on the circle, the theory becomes confining even in supersymmetric frameworks \cite{witten}.  
Indeed, with anti-periodic fermions and periodic bosons (which we shall assume from now on), at temperatures below $1/2\pi R$ the ground state becomes the so-called AdS Soliton \cite{witten,HM}, a horizon-less solution with zero entropy. For $T>1/2\pi R$ the BB (with its order $\NN^2$ entropy) dominates, allowing for a dual interpretation as a (de)confinement transition \cite{witten}.
Additionally, the Soliton exhibits certain quantum 
effects.~{\em E.g.}~its negative energy relative 
to AdS is dual to the CFT Casimir energy  \cite{HM}.

In \cite{NRT} (see also \cite{HW}), the `electrical'
response of the Soliton was studied by introducing a gauge field and a charged scalar. 
It was concluded that i) 
the confining/deconfining phases are  insulating/conducting,
the transition between them occurring
at $\mu$ and/or $T$ around $1/R$;
ii) both phases can exhibit SC bahaviour: 
below $T\sim 1/R$ and increasing $\mu$, one finds first a confined (Soliton) SC and then (for $\mu\gtrsim1/R$) a deconfined (BB) SC. 
Since the BB SC exhibits a conformal bahaviour, 
the SC/normal critical chemical potential is  $\mu_c\sim T/g$ for large $g$ 
(the charge of $\OO$)  
\cite{NRT}. 
By conformality also, the zero-temperature coherence length must  be  $\xi_0\sim 1/g\mu < 1/g\mu_c < R$
in the BB SC. 
In the Soliton SC, instead, conformality is broken in the infrared. There is a mass gap of order $1/R$
so one expects $\xi_0$ to be of order $R$ (as confirmed numerically).
From the previous discussion, then, one expects 
larger AB-effects (and the uplifting of LP degeneracy) 
in the Soliton SC   
than in the BB SC. In the Holographic setup,  
we will find an interesting additional suppression in the BB phase.

{\em Model.}
%
In the gravity picture of the HSC, the gravitational degrees of freedom are coupled to a $U(1)$ gauge field $A_\alpha$ and a complex scalar $\Phi$ according to the 
action 
$$
  S = \int d^{4} x \sqrt{- G}
  \{ (\mathcal{R} - \Lambda )/16 \pi G_N
  - \mathcal{F}_{\alpha \beta}^2/4
  - |D_{\alpha} \Phi |^2/L^2
  \},
$$
where 
$G_N$ is the Newton constant, 
$G$ is the determinant of the  spacetime metric $G_{\alpha\beta}$,
the cosmological constant $\Lambda$ 
determines the  AdS radius $L$ as $\Lambda=-{6/L^2}$  
and we omit a potential for the scalar for simplicity.
We use coordinates $(t,z,\chi,y)$ with $z$ the holographic direction such that the AdS-boundary sits at $z=0$. We consider finite temperature solutions  
with a compact spatial direction ($\chi$) of asymptotic radius $R$, 
and a noncompact one $y$. 
We also assume the `probe limit'
($g \rightarrow \infty$ with $g\mu$, $gA_\chi$ and $g\Phi$ fixed) 
where the gravitational backreaction of  $A_\alpha$, $\Phi$ is negligible \cite{HHH,NRT,HW},
so $A_\alpha$, $\Phi$ become the only dynamical fields on a fixed spacetime.
The relevant background metrics then are
$$
  ds^2 = (L^2/z^2)
  \left[f_t(z) dt^2 -dz^2/f(z) - f_\chi(z) d \chi^2 - dy^2 \right] \nonumber
$$
with $\{f_t,f_\chi\}=\{f(z),1\}$ ($\{1,f(z)\}$) 
for the neutral BB  
(AdS Soliton) 
where $f(z) = 1 - ( z/z_0)^3$ with $z_0=3/4\pi T$ ($3R/2$).
To study superconductivity in a fluxoid~$m$~we assume an ansatz 
$\Phi=\psi(z)\, e^{i m\chi/R}$, $A_\chi=A_\chi(z)$, $A_t=A_t(z)$ 
and solve numerically the equations of motion. 

\begin{figure}[t]
	 \includegraphics[width=7.5cm,height=3.2cm]{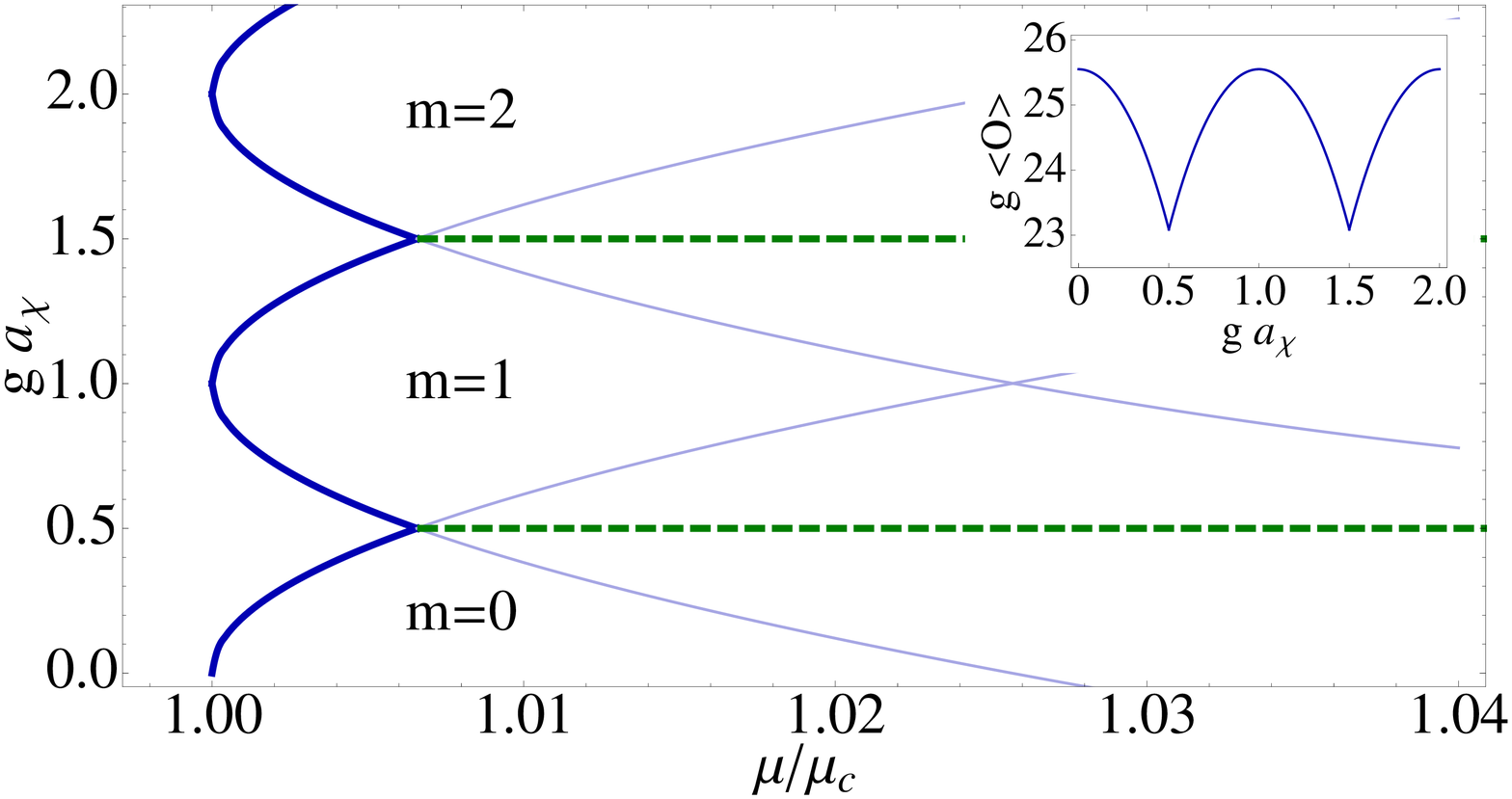}\\[-3mm]
  \caption{{\footnotesize Phase diagram for the BB SC at $T=1/\pi R$, $g\mu_cR=10.1$, 
  manifestly showing a LP effect.
  Thick solid blue lines separate the SC and normal phases. 
  Dashed green lines are transitions between different fluxoid sectors.
  Thin solid blue lines are existence lines for different fluxoid condensates.
  Inset: behaviour of the order parameter $\vev{\OO}$ for $\mu=1.03\mu_c$. Presented in units of $R$.
   }}\label{BH}
   ~\\[-8mm]
\end{figure}

The standard dual CFT interpretation 
starts from 
expressing the asymptotic behaviour of the gauge field  near $z=0$
as $A_\mu \rightarrow a_\mu+J_\mu z$.
One then identifies $J_\mu$ as the conserved $U(1)$ current carried by the CFT and 
$a_\mu$ as the conjugate `external' gauge potential which couples  to it. 
Similarly, from the boundary behavior  $|\Phi| \rightarrow c+\vev{\OO} z^3/3$
one identifies $\OO$ as  the condensing operator that breaks the $U(1)$ symmetry, and  
$c$ as its source term  (which we set to zero to realize  spontaneous symmetry breaking). 
Since we want to study the response to an external Wilson line  $a_\chi$ and to a chemical potential 
$\mu\equiv a_t$, the other boundary conditions
at $z=0$ are $A_{t} =\mu$ and $A_{\chi} =a_\chi$.

{\em Holographic LP Degeneracy.}
We start  with the BB background ($T>1/2\pi R$).
At the BB horizon ($z=z_0$) we have to impose the regularity conditions \cite{HHH,emergent}
\beqa A_t&=& 3\psi'  + z_0(g A_\chi-{\ds{ m/ R}})^2 \psi 
  \nonumber \\ &=&3 z_0 A_\chi' + {2g\psi^2} (g A_\chi-{ m/ R})=0 \nonumber ,\eeqa 
primes standing for $z$-derivatives.
With these boundary conditions 
the numerical integration of the field equations is straightforward. 
Our results, summarized in Fig.~\ref{BH}, were obtained using the Comsol package. 
The phase diagram in the $\mu$-$a_\chi$ plane (specifically, the SC/normal critical line) 
is clearly periodic in $a_\chi$ with period $1/gR$, indeed a dual LP effect.
This is a consequence of
the equations of motion and boundary conditions 
depending only on {\em local} gauge-invariants such as $|D_\chi\Phi|$.
The effective theory for the order parameter $\OO$, then, exhibits no 
non-local dependence on $W$,$m$. 
In particular, the order parameter $\vev{\OO}$ is continuous in the transitions between different $m$ channels. 
In brief, the expected Aharonov-Bohm effects end up entirely suppressed  in the BB SC
\footnote{
The Aharonov-Bohm effects would vanish even for $R<\xi_0$
if one chose periodic fermions on the circle 
(since these are compatible only with the BB solution).}. 
Of course, a `no-hair' theorem lies behind this suppression.

We shall assume 
that the charge of $\Phi$ is an integer multiple of the smallest charge in the theory, 
$g=N e$. 
In the CFT dual one can think that $\OO$ is  a composite operator made of $N$ singlet fields 
with $U(1)$ charge $e$. 
Since these do not participate in the solutions, $e$ (or $N$) is a free parameter 
which solely determines the realized
$Z_N$  symmetry and the range of $a_\chi$. 
We leave this as an unspecified parameter, 
since the LP effect implies the
degeneracy of the $N$ fluxoid sectors, whichever is $N$.

\begin{figure}[t]
	 \includegraphics[width=7.5cm,height=3.2cm]{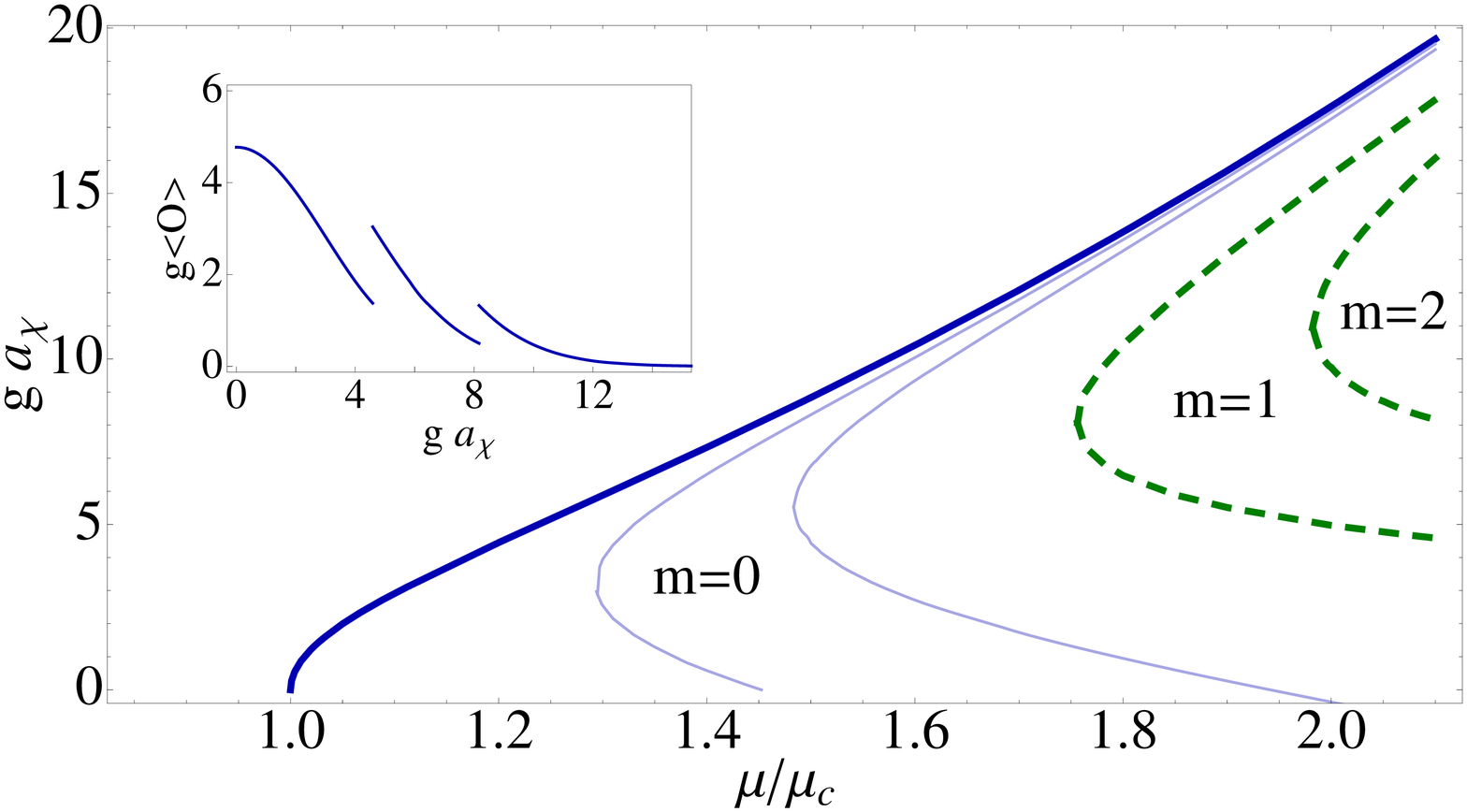}\\[-3mm]
  \caption{{\footnotesize Phase diagram for the Soliton SC, showing
  no LP degeneracy. 
  We have $g\mu_cR=1.81$.
  Line coding as in Fig.\ref{BH}.  
  Inset: form of $\vev{\OO}$ for $\mu=2.1\mu_c$. Presented in units of $R$.
   }}
   \label{Sol}
~\\[-7mm]
\end{figure}

{\em Holographic Lifting of the LP Degeneracy.}
We now turn to the Soliton  ($T<1/2\pi R$), 
which does display a Casimir energy \cite{HM} 
so it is expected to exhibit AB effects as well.
The only change with respect to the previous case is that there is now no horizon but instead the 
$\chi$ direction closes smoothly at $z=z_0$. 
The regularity conditions there are  \cite{soon}  $A_{\chi} =  0$, 
$3z_0 A_t' + 2 g^2A_t\psi^2 =0$ and
\be\label{regSol}
  0=3\psi'-g^2 z_0 A_t^2 \psi \; {\rm for~}m = 0 , \; \psi = 0\; {\rm for~} m \neq 0\,.
\ee
In addition, the same ansatz for $\Phi,\,A_\mu$ now leads to vortex solutions akin to those of 
\cite{vortex,emergent}
but having their core  at the infrared tip of the holographic direction $z$.

Our results for the numerical integration of the equations are summarized in Fig.~\ref{Sol}. As in the BB phase, for large enough $a_\chi$ the condensation is favored in the $m\neq0$ 
sectors
\footnote{
Multi-center solutions could have lower energy than the single vortex for $|m|>1$. The $m=2$ lines in Fig.~\ref{Sol}, then, represent `upper' bounds on the actual lines.
}. 
However, now the domains occupied by different sectors do not appear regularly -- the $1/gR$ periodicity is completely lost. 
%
Still, one can view the phase diagram as a (large) deformation of the BB one,
and it can be reproduced by an effective theory 
whose coefficients have non-local $W$,$m$-dependence:
AB effects are indeed present.   
In the gravity theory, this 
arises through the boundary conditions \eqref{regSol}, which
depend directly on 
$m$ and $A_\chi$.
Let us also stress that the order parameter $\vev{\OO}$ exhibits jumps in the transitions between different $m$ sectors, a manifestation of the coupling $\LL_{j-\theta}$.

%

{\em Little-Parks and Quantum hair.}
HSCs suggest an interplay between BH physics and superconductivity.
The mere existence of the BB SC is a consequence of the fact that the AdS space evades the no-hair (or uniqueness) theorems: not only regular AdS BHs with scalar hair exist, they even become the ground state at low temperatures. 
Another well known departure from the spirit of the no-hair theorems is given by the so-called Quantum Hair (QH) -- the `charges' which are measurable quantum-mechanically but not classically. 
A sharp example  is given by the discrete gauge symmetries discussed above: 
the charge defined by the
unbroken $Z_N$ gauge group  is measurable at long distances, yet it is associated to a massive gauge field. 
Then, BHs must be able to carry this charge 
without classically affecting  the spacetime  \cite{KW,CPW}.
Note that even if this extends the kinds of hairs supported by BHs, these still comply with a 
{\em classical-uniqueness} theorem: classically, BH solutions are insensitive to the amount of QH (see footnote 22).

The connection between quantum hair and the LP effect becomes apparent once one realizes that the defining property of QH --classical undetectability-- 
implies the (classical) {\em degeneracy} 
between different discrete charge sectors. 
This suggests that the LP effect 
results from a magnetic form of discrete charge acting as QH.
To be more precise,
let us split the spatial gauge field as $a_\chi \equiv m' /gR + \wt a_\chi$ with $m'$ an integer and $\wt a_\chi$ the non-integer part of $a_\chi$ {\em modulo} ${1/gR}$. 
In the BH phase all quantities depend only on 
$\wt a_\chi$ 
and on $m-m'$. Therefore for every choice of $\wt a_\chi$,
the configurations $m=m'=k$ with $k=0,... N-1$ are  degenerate classically. These configurations are a magnetic counterpart of discrete gauge charge: there is an $N$-fold of them and the  winding number $m$ is locked to the  `magnetic flux' $m'$. %
%
Since this is a discrete gauge charge it is a QH, and classically it must leave the BB solutions unaffected even including the backreaction (since both the stress tensor and boundary conditions depend only on local operators).  
Hence, in the BB SC 
the different magnetic sectors
are classically degenerate, 
and the short `LP' period $\Delta a_\chi = 1/gR$ follows. 
Instead, the Soliton SC  
is horizon-free and does not obey  classical-uniqueness. 
Indeed, {\em the Soliton is  classically sensitive to QH}:
the closing of the spatial circle 
at $z=z_0$ enforces 
Wilson line-dependent boundary conditions and yields the fundamental $1/eR$ periodicity
\footnote{
It seems that a `proof' of {\em classical uniqueness} 
in our context follows by noting that
i) classical sensitivity to QH can arise only from a collapsing spatial circle (CSC);
and ii) no {\em regular} BB solutions with CSCs exist.}.

The discussion so far treated AdS gravity classically, which is valid for small curvatures. 
Quantum corrections in the gravity theory 
will generically spoil the LP periodicity, even if by a small amount. 
Again, the simplest quantum effect is the AB effect from 
fundamental charges winding around the circle, 
which exhibits the fundamental period $1/eR$ 
and need not be exponentially suppressed
\footnote{Uplifting of the {\em electric} QH 
proceeds via AB-inter\-fe\-rence of virtual magnetic flux tubes, 
so it is exponentially suppressed \cite{CPW}.
The instantons that capture this effect 
are almost identical to the vortices on the Soliton.}.
According to the AdS/CFT dictionary, quantum effects in the gravity side map to
$1/\NN$ corrections in the large $\NN$ expansion of the CFT. 
One infers that the AB effect is  $1/\NN$-suppressed  in the CFT deconfined 
plasma phase.
By contrast, the classical sensitivity to QH
exhibited by the Soliton 
means that AB effects are
{\em un}suppressed in the confining phase.
In CFT terms, this can be understood
from  i)~the large $\NN$ is a classical limit 
\cite{Yaffe}
and ii) the deconfined plasma state should have a classical counterpart 
(in which quantum effects do die out with $1/\NN$)
while the confining state should not.
According to this interpretation, the BH classical uniqueness theorem 
implies a dual `theorem': the deconfined plasma states  have a classical limit at large $\NN$.

As a final remark, one may speculate that
the properties  found here for the (de)confining SCs 
may extend beyond the applicability of the large $\NN$ limit.
If so, it is tempting to infer that the lifting of the LP effect
should be more (un)suppressed  in (un)conventional SCs.

\paragraph*{Acknowledgments}
We thank G Dvali, A Pomarol, A Sanchez and S Sibiryakov for useful discussions. 
This work was supported by UAB, 
the EU ITN (PITN-GA-2009-237920) and by MIUR (2006022501).
\vspace{-0.6cm}

\end{document}